\documentclass{moriond}

\def\Journal#1#2#3#4{{#1} {\bf #2}, #3 (#4)}

\def\PLB{{\em Phys. Lett.}  B}

\def\be{\begin{equation}}
\def\ee{\end{equation}}
\def\bea{\begin{eqnarray}}
\def\eea{\end{eqnarray}}

\newcommand{\Photo}{\includegraphics[height=35mm]{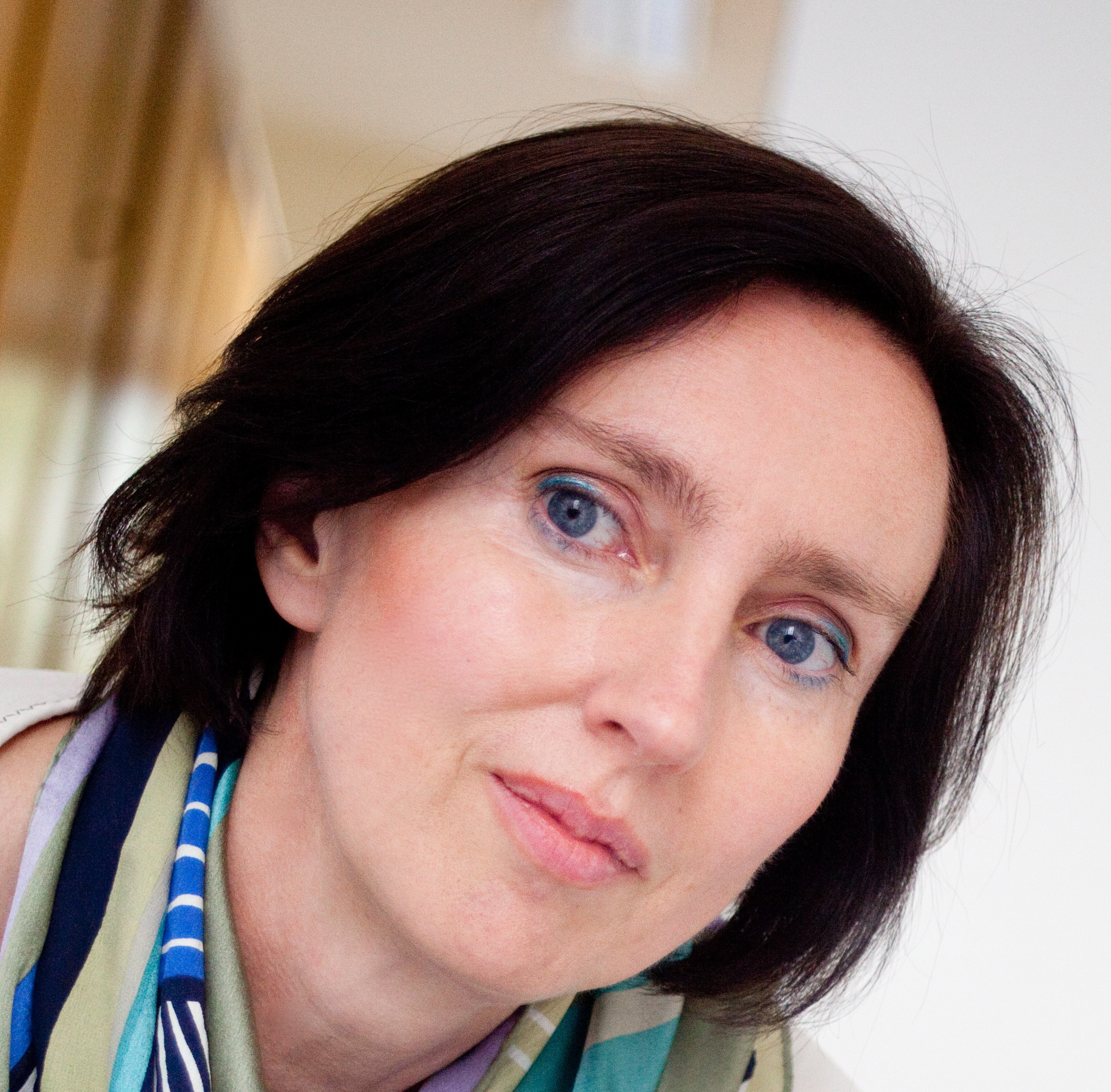}}

\begin{document}
\vspace*{4cm}
\title{Memories from the W Boson Discovery}

\author{ Claudia-Elisabeth Wulz }

\address{Institute of High Energy Physics of the Austrian Academy of Sciences,\\
Nikolsdorfergasse 18, 1050 Vienna, Austria}

\maketitle\abstracts{
The fascinating story of a major discovery at CERN is outlined. The bold decision to convert its most powerful, and only recently inaugurated, proton accelerator to a proton-antiproton collider led to the discovery of the W and Z bosons -- mediators of the weak interaction -- in a record time, at the experiments UA1 and UA2. The decisive roles of Carlo Rubbia and Simon van der Meer, who received the 1984 Nobel Prize for physics, are underlined. 
}

\section{Introduction}
About 40 years ago, the time was ripe for a crucial discovery to confirm a cornerstone of the standard model of particle physics. Direct evidence for neutral currents had already been found with neutrino interactions recorded at the Gargamelle bubble chamber at CERN in 1973 \cite{NeutralCurrents}. The measurement of cross-section ratios between neutral and charged current interactions enabled the setting of limits for the Weinberg weak mixing angle, $\theta_W$, which is in turn related to the mass of the W boson, $M_W$, through the formula
\begin{equation}
M_W = \sqrt{\frac{\pi \alpha}{\sqrt 2 G_F}} \frac{1}{\sin \theta_W} \approx \frac{37~\mathrm{GeV}}{\sin \theta_W}. 
\end{equation}
By the end of the 1970's $M_W$ was thus predicted to be approximately in the range 60 to 80 GeV. CERN's Super Proton Synchrotron (SPS) had just begun its single-beam operation in 1976. In the same year, Rubbia, McIntyre, and Cline proposed to transform a conventional accelerator to a proton-antiproton collider \cite{RubbiaMcIntyreCline}, in order to be able to reach the energy threshold and the necessary luminosity to produce  W and Z events. The rest is history. The decision to convert the SPS into the Super Proton-Antiproton Synchrotron (Sp$\overline\mathrm{p}$S) was made in 1978. Within only three years, the first collisions were recorded at a centre-of-mass energy $\sqrt s = 540~\mathrm {GeV}$. By the end of 1982, enough data had been accumulated to enable the discovery of the W, and its announcement came in 1983. 

\section{The Super Proton-Antiproton Synchrotron and the experiments}
\begin{figure} [hbt]
\centerline{\includegraphics[width=0.7\linewidth]{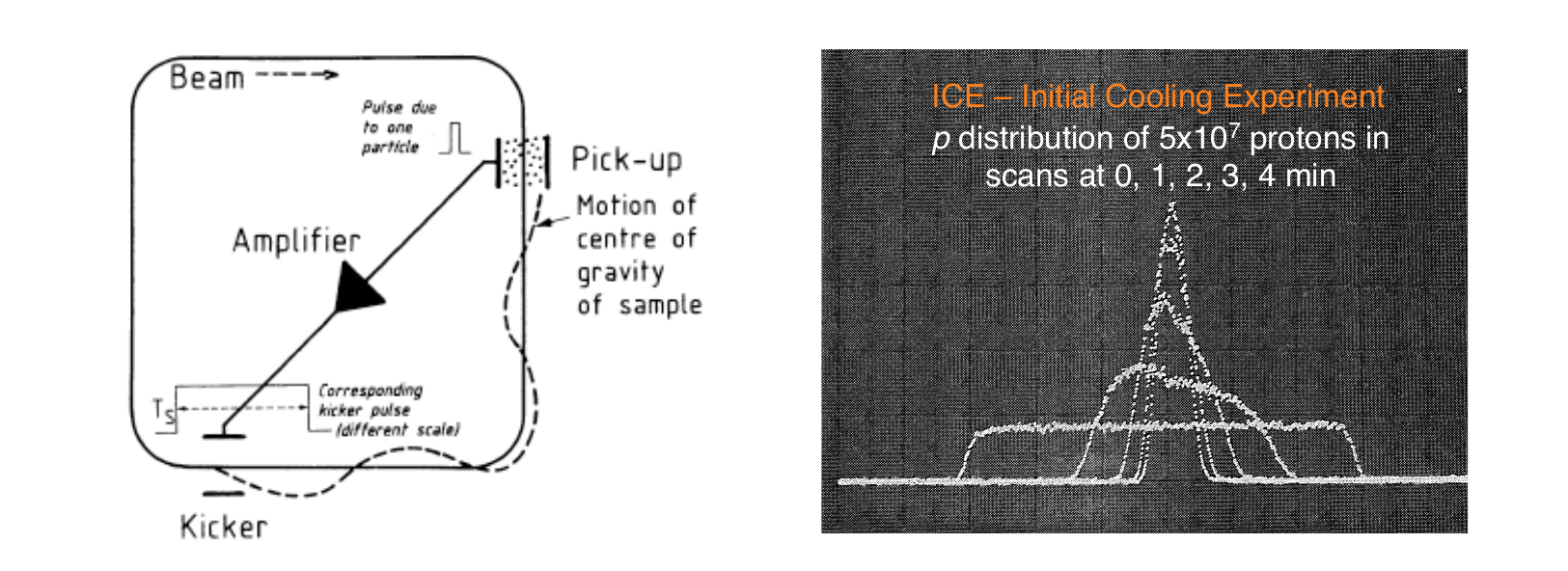}}
\caption[]{Principle of stochastic cooling for horizontal oscillations (left) \cite{Moehl}, Initial Cooling Experiment (right) \cite{Koziol}}
\label{fig:StochasticCooling}
\end{figure}

The transformation of the SPS into the Sp$\overline\mathrm{p}$S was no small feat. It relied on a key technique to produce and store dense beams of protons and antiprotons -- stochastic cooling, invented by Simon van der Meer \cite{VdM}. It reduces the energy spread and angular dispersion of particle beams. The principle is illustrated in Fig.~\ref{fig:StochasticCooling} for horizontal oscillations around the nominal orbit. A pick-up electrode samples the distance of the centre-of-gravity of a group of particles from this orbit. The corresponding signal is amplified and sent through an approximately diagonal line to a kicker. The signal path is shorter than the particle path, and the kicker is thus able to apply an electric field when the particles pass in order to correct the deviation. The Initial Cooling Experiment (ICE) demonstrated that momentum cooling could indeed be achieved. Out of 240 antiprotons that were cooled, 80 still circulated four days later. This convinced CERN to go ahead with the Sp$\overline\mathrm{p}$S project, and the associated construction of the Antiproton Accumulator \cite{Koziol}.

\begin{figure} [hbt]
\centerline{\includegraphics[width=0.9\linewidth]{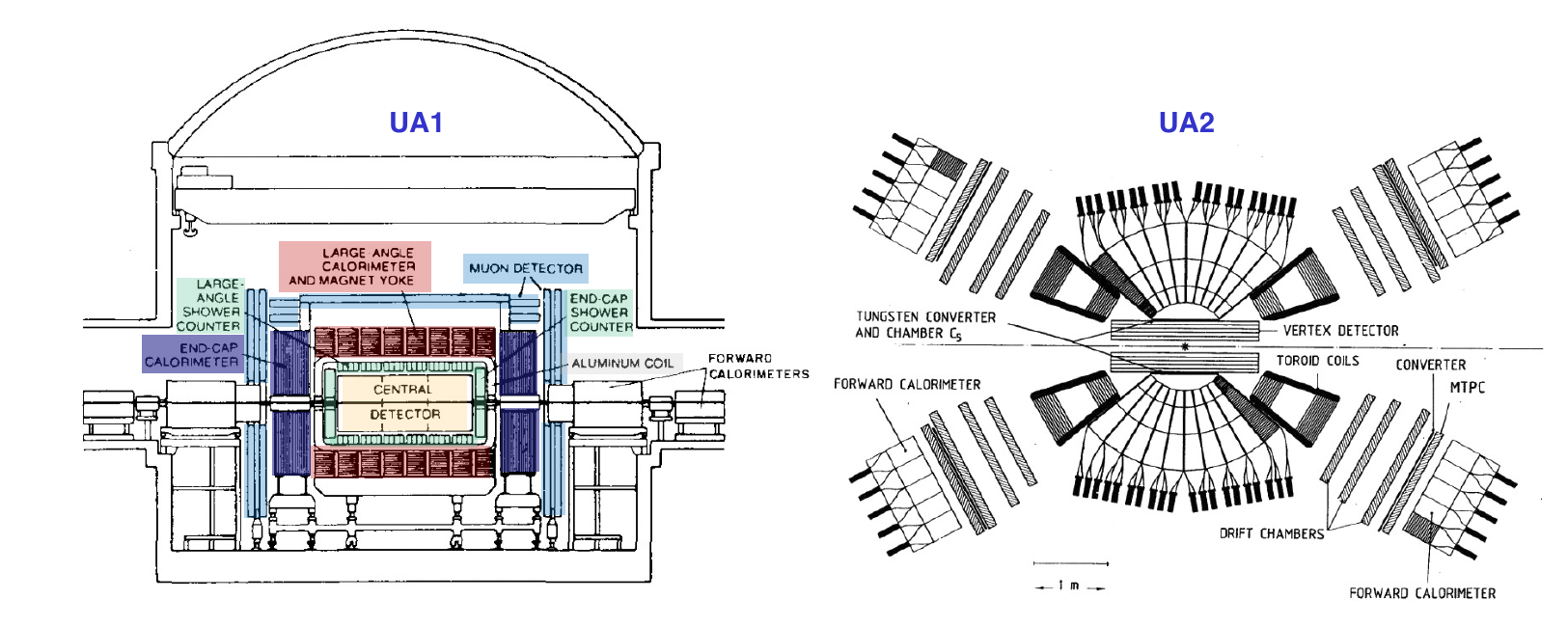}}
\caption[]{UA1 Experiment \cite{UA1proposal} (left), UA2 Experiment \cite{UA2proposal} (right)}
\label{fig:UA1UA2}
\end{figure}

Two experiments were conceived and built within a very short time, in underground areas along the Sp$\overline\mathrm{p}$S tunnel. UA1 \cite{UA1proposal} and UA2 \cite{UA2proposal} were both approved in 1978, and started data taking in 1981. While UA1 was the first ``hermetic" multi-purpose detector, UA2 was more limited in scope, particularly because it did not have a muon system. Both experiments were designed to detect electrons, photons, hadrons, and neutrinos, using the ``missing energy" technique for their identification. UA1 alse detected muons. Figure \ref{fig:UA1UA2} shows the setup of the two detectors.

UA1's central detector was the largest drift chamber of its time. It was surrounded by a lead-scintillator electromagnetic calorimeter. A warm dipole magnet provided a flux density of 0.7 Tesla. The instrumented return yoke served as a hadron calorimeter, made of iron-scintillator plates. Large muon chambers and forward calorimetry up to 0.2$^0$ from the beam line completed the experiment. UA2 was optimized for the detection of electrons from W and Z decays, but was also well suited to measure jets. Its vertex detector was made of cylindrical tracking chambers, followed by a preshower detector consisting of a tungsten converter and a multi-wire proportional chamber for electron identification, and a highly-granular calorimeter, made of a lead-scintillator electromagnetic and an iron-scintillator hadronic section. The central calorimetry had a spherically projective geometry, and was complemented by forward calorimetry, which initially covered a region up to 20$^0$ from the beam axis. It was extended to 5$^0$ by the end of 1985. Toroid coils provided a magnetic field in the forward regions, where the W decay asymmetry is maximal. 

\section{The path to the discovery of the W boson}
\begin{figure} [hbt]
\centerline{\includegraphics[width=0.8\linewidth]{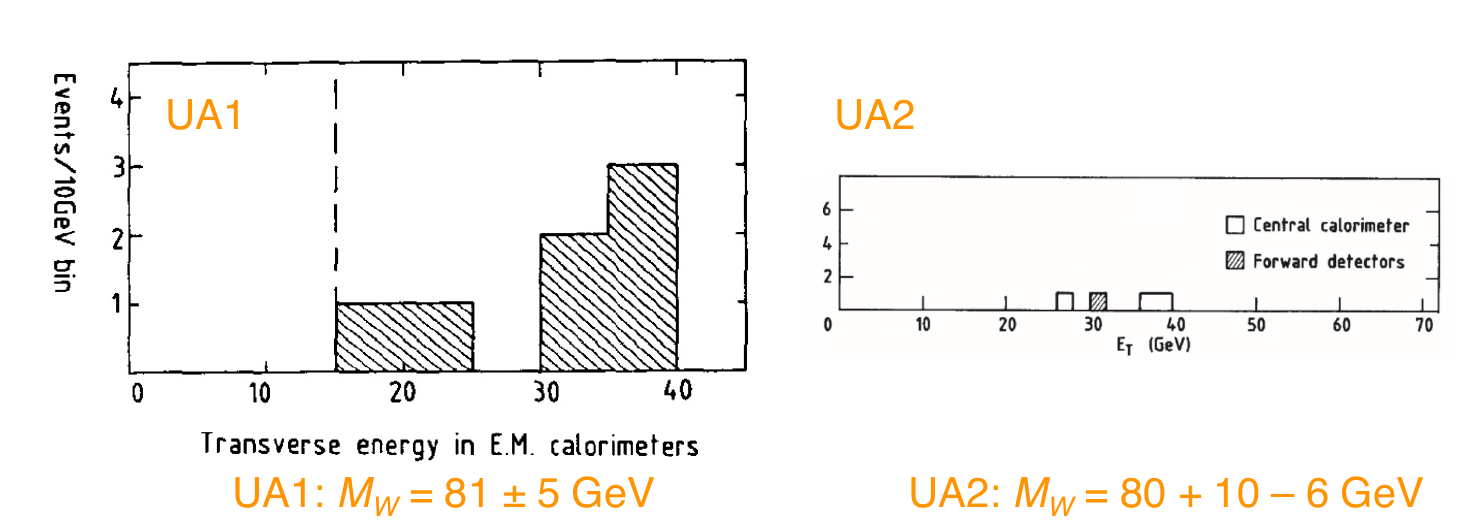}}
\caption[]{W mass measurement in UA1 \cite{WdiscoveryUA1} (left), W mass measurement in UA2 \cite{WdiscoveryUA2} (right)}
\label{FIG:Wmass1983}
\end{figure}

\begin{figure} [hbt]
\centerline{\includegraphics[width=0.8\linewidth]{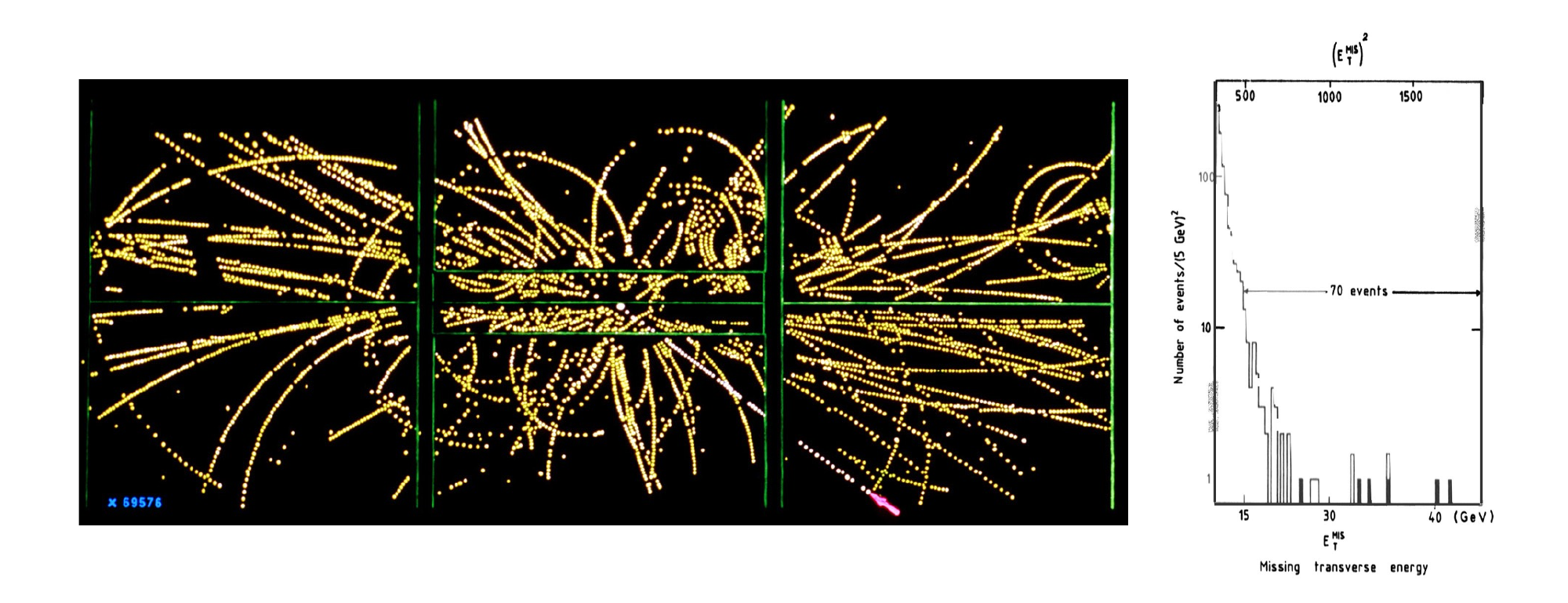}}
\caption[]{UA1 W event \cite{WeventCDS} (left), and missing transverse energy distribution \cite{WdiscoveryUA1} (right)}
\label{fig:Wevent}
\end{figure}

The first physics data came in 1981, with UA1 recording collisions already in July, and UA2 in December of that year. The latter data taking period was referred to as the ``jet run", and resulted in about 20~$\mu$b$^{-1}$ of integrated luminosity \cite{DenegriCERNCourier}. During this time UA1 focused more on the tracker, whereas UA2 concentrated on calorimetry. UA2 thus presented better results on back-to-back 2-jet events at the 21$^\mathrm{st}$ International High Energy Physics Conference in Paris in Aug. 1982 \cite{Repellin}. In early Nov.~1982, a W $\rightarrow e \nu_e$ candidate with an isolated electron and missing energy was found in UA1, but there was some hadronic activity. A few days later, a ``gold-plated" candidate was found. By the end of 1982, with 28~nb$^{-1}$ integrated luminosity achieved, 6 W events were found in UA1 \cite{WdiscoveryUA1}, and 4 in UA2 \cite{WdiscoveryUA2}. The characteristic properties per event were an isolated electron with high transverse momentum, a large amount of missing energy, and a Jacobean peak at $M_W/2$ in the distribution of the electron transverse momentum of all events. In UA1, two search methods were used, applying a method with a stringent electron selection on one hand, and another one called the ``Saclay missing energy method" \cite{DenegriCERNCourier}. Both methods led to the same 5 events, and eventually 6 events from the first method made it to the UA1 W discovery paper. It should be noted that only the W decay channel into electrons contributed to the discovery. The events leading to the announcement of the discovery were the following. On 12 Jan.~1983, Carlo Rubbia and Pierre Darriulat made presentations at the Third Topical Workshop on p$\overline\mathrm{p}$ Collider Physics in Rome, entitled ``Jets, large p${_T}$, etc." and ``Preliminary searches for hadron jets and for large transverse momentum electrons at the SPS p$\overline\mathrm{p}$ collider", respectively. These titles were obviously intentionally cryptic, but the seminars given at CERN by Carlo Rubbia and Luigi Di Lella on 20 and 21 Jan.~1983, respectively, left no doubt that the discovery of the W boson was now history. The CERN press conference of 25 Jan.~1983 contained the statement ``This is indeed a major step forward in contemporary physics". The mass of the W boson was measured as depicted in Fig.~\ref{FIG:Wmass1983}. 
UA1 quoted a value of $M_W = 81 \pm5$~GeV, whereas UA2 quoted $M_W = 80 + 10 - 6$ GeV, where the latter uncertainties refer to statistical and systematic uncertainties, respectively. 
A beautiful W event recorded by UA1 is shown in Fig.~\ref{fig:Wevent}. The stiff electron track is indicated by the white trajectory and pink arrow. The missing transverse energy distribution for W signal (black boxes) and background events is also shown. 

\section{Conclusions, outlook, and acknowledgments}
The 1984 Nobel Prize for Physics was awarded to Carlo Rubbia and Simon van der Meer
``for their decisive contributions to the large project which led to the discovery of the field particles W and Z, communicators of weak interaction”. I had the privilege to be part of this adventure, and am most grateful to Carlo for welcoming me to UA1 in 1982 already as a summer student, and later as a fellow. His charisma and courage to strive for the impossible have been inspiring to me, and crucial to bring this endeavour to a more than successful conclusion. He anticipated it by saying "Se sono rose, fioriranno" (referring to the W events: ``If they are roses, they will flower") at the Rome conference. Since then, precise measurements of the W mass have been underway. Recently the ATLAS Collaboration has published a measurement that is compatible with the standard model \cite{ATLASWmass2023}, but there are also contradictory results by the CDF Collaboration~\cite{CDF}, which have sparked interest. The future will tell us if surprises are in store.  

I would like to thank the organizers for a wonderful conference, as always, and in particular Boaz Klima, for inviting me to give this presentation.

\end{document}